\documentclass[prb,aps,showpacs,preprint]{revtex4}

\usepackage{epsfig}
\usepackage{graphicx}

\begin{document}


\title{Evidence for s-wave pairing from  measurement on lower critical field  in  $MgCNi_3$}
\author{X. F. Lu, L. Shan, Z. Wang, H. Gao, Z. A. Ren, G. C. Che}
\author{H. H. Wen\footnote{Mail of the corresponding author: hhwen@aphy.iphy.ac.cn}}\email{hhwen@aphy.iphy.ac.cn}

\affiliation{National Laboratory for Superconductivity, Institute
of Physics, Chinese Academy of Sciences, P.~O.~Box 603, Beijing
100080, P.~R.~China}

\begin{abstract}
Magnetization measurements in the low field region have been
carefully performed on a well-shaped cylindrical and an
ellipsoidal sample of superconductor $MgCNi_3$. Data from both
samples show almost the same  results. The lower critical field
$H_{c1}$ and the London penetration depth $\lambda$ are thus
derived. It is found that the result of normalized superfluid
density $\lambda^2(0)/\lambda^2(T)$ of $MgCNi_3$ can be  well
described by BCS prediction  with the expectation for an isotropic
s-wave superconductivity.
\end{abstract}

\pacs{74.25.Bt, 74.25.Ha, 74.70.Dd}

\maketitle

\section{INTRODUCTION}

The pairing symmetry  is very essential for uncovering the
mechanism   both for conventional and high-$T_c$
superconductivity. The recently discovered intermetallic
perovskite $MgCNi_3$ superconductor\cite{he-nature} is regarded as
a bridge between conventional superconductors and high-$T_c$
cuprates, and the issue concerning its symmetry of order parameter
has attracted considerable  attention. However, pairing symmetry
about $MgCNi_3$ remains highly controversial in reported
literatures. NMR\cite{singer-NMR}, specific heat\cite{lin-SH},
scanning tunneling measurement\cite{kinoda-scanning} and point
contact tunneling spectra\cite{shan-tunneling} favor the s-wave
pairing in $MgCNi_3$. On the other hand, the earlier theoretical
calculation\cite{rosner-prl}, the tunneling
spectra\cite{mao-tunneling} and the penetration depth
measurement\cite{prozorov-penetration} support non-s-wave
superconductivity. Recently a two-band s-wave model has been
proposed by W$\ddot{a}$lte \textsl{et al.}\cite{walte} who try to
explain the complex behavior observed in $MgCNi_3$.

In this paper, we  derive the  thermodynamic parameters $H_{c1}$
and $\lambda$ of two $MgCNi_3$ samples by careful magnetization
measurement. It is found that the normalized superfluid density,
$\lambda^{2}(0)/\lambda^{2}(T)$, can be described by  BCS
prediction for a s-wave pairing symmetry. Therefore, our
magnetization data support the conventional single band s-wave
superconductivity in $MgCNi_3$.

This paper is organized as follows: The samples and experimental
details  are presented in  section II. The data and  discussions
are given in  section III. And  section IV gives the summary.

\section{SAMPLES AND EXPERIMENTAL DETAILS}

The polycrystalline $MgCNi_3$ sample investigated here has been
prepared by powder metallurgy method, and  the details of
preparation  can be found elsewhere\cite{ren-sample}. The
superconducting transition temperature is 6.9 K measured by both
magnetization [ ac susceptibility ( $f=333$ Hz, $H_{ac}=1$ Oe )
and dc diamagnetization shown in Fig. 1(a) ] and resistivity
measurement. The $M(T)$ curves show a sharp transition with the
transition width less than 0.5 K. The x-ray diffraction ( XRD )
analysis presented in Fig. 1 (b) shows that all diffraction peaks
are from the $MgCNi_3$ phase, which indicates that the sample is
nearly of single phase.

\begin{figure}
       \centering
       \includegraphics[width=9 cm] {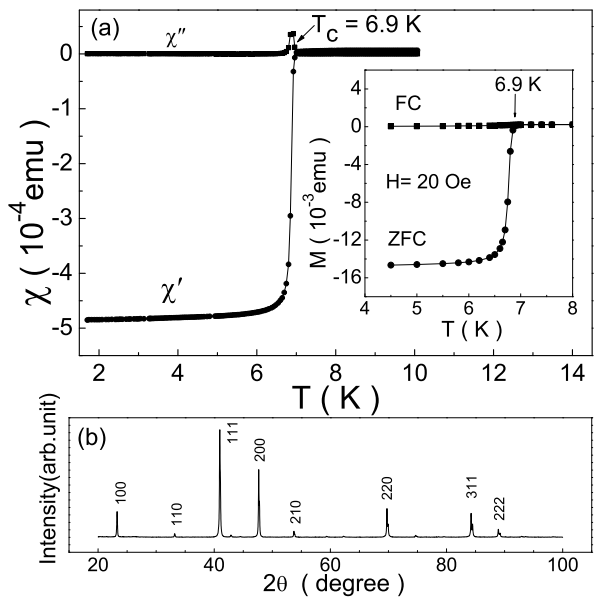}
       \caption{(a) The ac susceptibility  curves of $MgCNi_3$ measured by MagLab with ac field 1 Oe and frequency 333 Hz.
       The inset gives the ZFC and FC  dc diamagnetization at 20 Oe measured by SQUID. (b) XRD patterns of the $MgCNi_3$.}

       \label{figure1}
\end{figure}

In order to minimize the demagnetization factor, one sample (
denoted as $S$-$c$ ) has been carefully cut and ground to a
cylinder with a diameter of 1.1 mm and length of 7.0 mm. The
demagnetization factor in this situation is almost negligible
since the field has been applied along the axis of the cylinder.
Another sample ( denoted as $S$-$e$ ) has been polished to an
ellipsoid with semi-major axis a=3.74 mm and semi-minor axis b=1.5
mm. The demagnetization factor for the ellipsoidal sample is
$n=(1-\frac{1}{e^2})(1-\frac{1}{2e}\cdot
\ln\frac{1+e}{1-e})\approx 0.136$, with $e=\sqrt{1-b^2/a^2}$. The
magnetic fields have been applied parallel to the longitudinal
axis of the samples.

The magnetic measurements are mainly carried out on an Oxford
cryogenic MagLab system ( MagLab12Exa, with temperature down to
1.5 $K$ ) and checked by a quantum design superconducting
interference device ( SQUID, MPMS 5.5 T ). After zero-field cooled
( ZFC ) from 25 K to a desired temperature, the magnetization
curve $M(H)$ is measured with the applied magnetic field swept
slowly up to 1000 Oe ( $\gg$ $H_{c1}$ ). It is important to note
that the magnet has been degaussed at T= 25 K in order to
eliminate the remanent field before each measurement. It is
essential to do degaussing since otherwise even 5 Oe residual
field may cause significant effect on the result of magnetization.

\section{EXPERIMENTAL DATA AND DISCUSSIONS}
In this section, the  processes to obtain the superconducting
parameters by magnetization measurement have been reported in
detail for two $MgCNi_3$ samples, one  is a cylinder and another
is an ellipsoid.

\subsection{The cylindrical sample ( $S$-$c$ )}

The curves of dc magnetization  are shown in Fig. 2. The
temperature varies between  1.57 K and 6.88 K with steps  0.2 K (
some 0.4 K ). All curves show clearly the common linear dependence
of the magnetization on field caused by Meissner effect at low
fields, and this extrapolated common line is the so-called
\textquotedblleft Meissner line\textquotedblright ( ML ). The
optimal ML ( solid line in Fig.2 ) is achieved by doing linear fit
$M(H)$ of the lowest temperature ( 1.57 K ) at low fields, which
represents the magnetization curve of Meissner state. The value of
H$_{c1}$ is determined by examining the point of departure from
linearity on the initial slope of the magnetization curve ( ML )
with a certain criterion. The results of subtracting this ML from
magnetization curves are plotted in Fig. 3 and the $\Delta M$
between $7.0\times 10^{-4}$ and $1.4\times10^{-3}$ emu are shown
in the inset with an enlarged view. All curves show a fast drop to
the resolution of device when the real $H_{c1}$ is approached, so
the value of H$_{c1}$ is easily obtained by choosing a proper
criterion of $\Delta$M. The H$_{c1}(T)$ acquired by using criteria
of $\Delta M = 7.0 \times 10^{-4}$ and $1.1 \times 10^{-3}$ emu
are shown in Fig. 4. Then the penetration depth $\lambda(T)$ can
be achieved from H$_{c1}(T)$ by
\begin{equation}
H_{c1}=\frac{\rm \Phi_{0}}{\rm 4\pi \lambda ^2}\ln\kappa
\label{eq1}
\end{equation}
and they are displayed in the inset of Fig. 4. Here
$\Phi_{0}=hc/2e\approx 2\times 10^{-7} G\cdot cm^2$ is the flux
quantum, and $\kappa$ is the Ginzburg-Landau parameter. We take
$\kappa$ as constant  since it is a weakly temperature dependent
parameter.

\begin{figure}
       \centering
       \includegraphics[width=8cm]{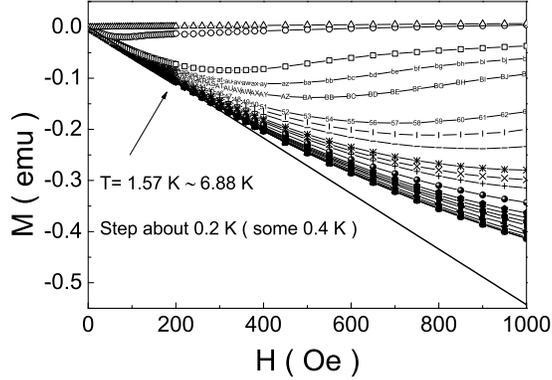}
       \caption{The magnetization curves of $M(H)$ for $MgCNi_3$. The solid line is the \textquotedblleft Meissner line\textquotedblright~defined in the text.
       The temperature is varied from 1.57 K to 6.88 K from  bottom  to  top,
       with  step about 0.2 K ( some 0.4 K ). It is found that the initial slope of all the curves is the same. }
       \label{figure2}
\end{figure}

\begin{figure}
       \centering
       \includegraphics[width=8cm]{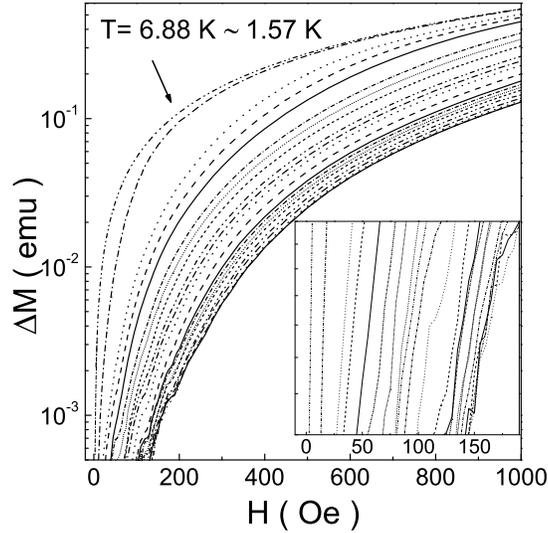}
       \caption{The difference between the \textquotedblleft Meissner line\textquotedblright~and the $M(H)$ curves. $\Delta M$ is shown in
       logarithmic scale. The inset shows the enlarged  $\Delta M$ ( between $7.0\times
       10^{-4}$ and $1.4\times 10^{-3}$ emu ).}
       \label{figure3}
\end{figure}

\begin{figure}
       \centering
       \includegraphics[width=8cm]{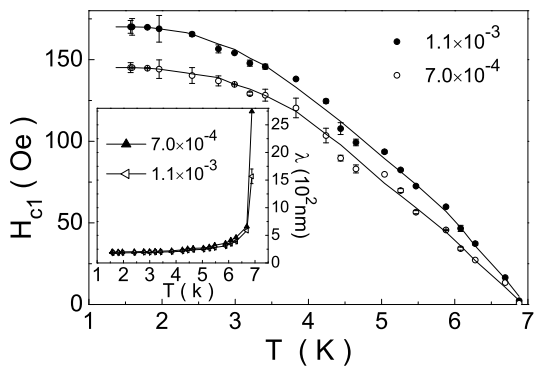}
       \caption{The temperature dependence of the  nominal $H_{c1}$ and $\lambda$ ( inset ).
       $H_{c1}(T)$ is obtained by using criterion of $\Delta M = 7.0\times 10^{-4}$ ( open circles )
       and  $1.1\times 10^{-3}$ emu ( solid circles ), respectively. Error bars are given for  determining the nominal
       $H_{c1}$ and  $\lambda$. Lines are guides to the eyes.}

       \label{figure4}
\end{figure}

\begin{figure}
       \centering
       \includegraphics[width=8cm]{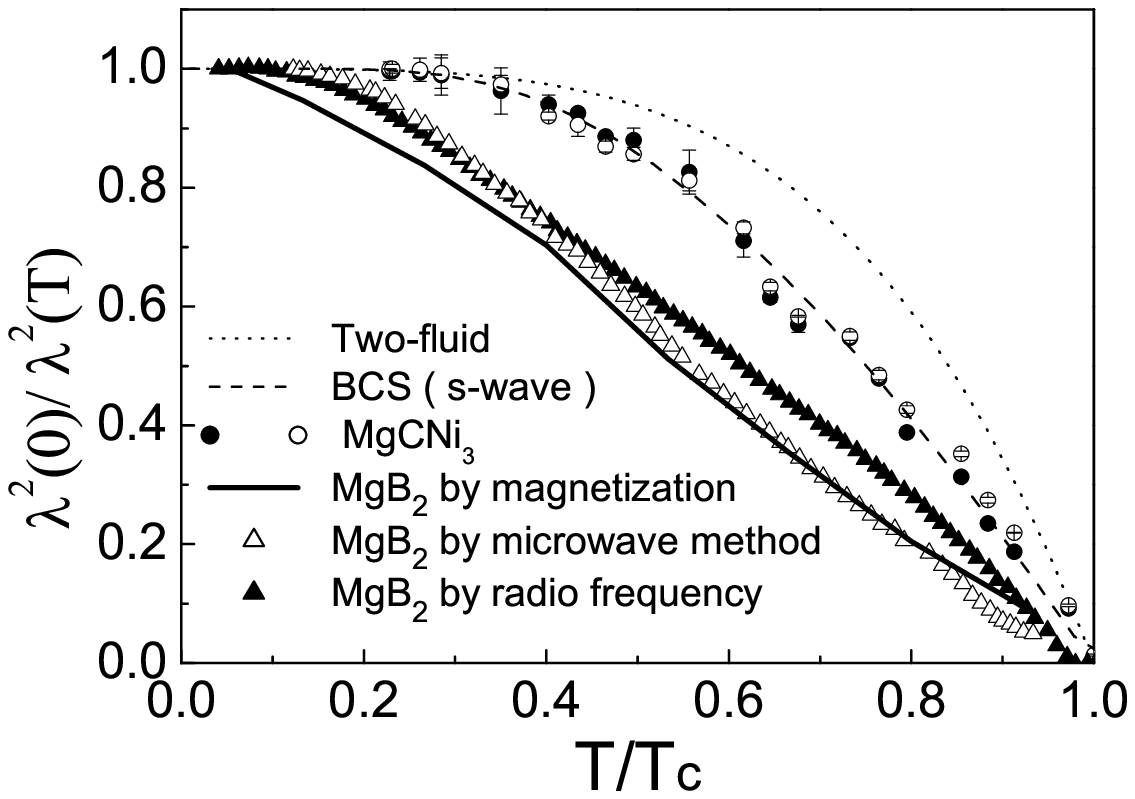}
       \caption{The temperature dependence of $\lambda^2(0)/\lambda^2(T)$ for $MgCNi_3$
       and $MgB_2$. The values of $\lambda^2(0)/\lambda^2(T)$ are
       obtained by $H_{c1}(T)$ using criteria of $\Delta M$ equal $7.0\times 10^{-4}$ ( open circles ) and
       $1.1\times 10^{-3}$ emu ( solid circles )  for $MgCNi_3$ and $1.0\times 10^{-4}$ emu for $MgB_2$( solid).
       The  prediction by two-fluid model ( dotted ) and  BCS s-wave(  dashed ) are also shown.
       The data of  open and solid triangles are the experimental
       measurement by microwave resonator method and radio
       frequency technique for $MgB_2$, respectively ( from Ref.[\onlinecite{klein-appliedSC}] and Ref.[\onlinecite{manzano-prl}] ).}
       \label{figure5}
\end{figure}

The values of nominal H$_{c1}$ and  $\lambda$ seem to be criterion
dependent in this method, however temperature dependence of
$H_{c1}$ and $\lambda (T)$ are found to be weekly criterion
dependent if the data is normalized by the zero temperature values
( see $\lambda^{2}(0)/\lambda^{2}(T)$ in Fig.5 ) . In addition,
$\Delta M$ drops sharply with decreasing magnetic field, the use
of lower $\Delta M$ value in our criterion will not result in a
much different $H_{c1}(T)$ curve. If not specially mentioned, the
discussion is based on the data using the criterion of $7.0\times
10^{-4}$ emu hereinafter. At the temperatures below 2.8 K ( $<$
0.4 $T_c$ ), the values of $H_{c1}(T)$ and $\lambda(T)$ are almost
constant despite the lack of the data below 1.5 K. This may imply
the conventional s-wave nature in $MgCNi_3$, because the finite
energy gap manifests itself with an exponentially activated
temperature dependence of thermodynamic parameters.  This can be
further confirmed in the following discussion on superfluid
density. Worthy of noting is that, it is very difficult to
distinguish a slight difference of $H_{c1}$ ( or 1/$\lambda^{2}$ )
in low temperature region between different pairing symmetries,
for example for an ideal s-wave, an exponential dependence is
anticipated, for a dirty d-wave, a quadratic form
$\rho_s(T)=\rho_s(0)-\alpha T^2$ is expected. Here we use an
alternative way, i.e, to fit the  data in intermediate and high
temperature region to extract useful message for pairing symmetry.

As we know, the total superfluid density $\rho_s$ is proportional
to $\lambda^{-2}(T)$, and $\lambda^{2}(0)/\lambda^{2}(T)$
represents the normalized superfluid density. In Fig. 5, we
display the temperature dependence of
$\lambda^{2}(0)/\lambda^{2}(T)$ of $MgCNi_3$ with $\lambda(0)$ as
a fit parameter. The  predictions of  BCS s-wave( dashed ) and
two-fluid model ( dotted  ) are also shown. According to the BCS
theory for clean superconductors\cite{Tinkham,kim-prb}, the
normalized superfluid density $\lambda^{2}(0)/\lambda^{2}(T)$ is
expressed as follows:
\begin{equation}
\frac{\lambda^{2}(0)}{\lambda^{2}(T)}=1-2\int^{\infty}_{\Delta(T)}{(-\frac{\partial{f(E)}
 }{\partial{E}}) D(E)} \mathrm{d} E
\end{equation}
where $\Delta(T)$ is the  BCS superconducting energy gap,
$f(E)=1/[exp(-E/{k_B}T)+1]$ is the Fermi distribution function,
and $D(E)=E/{(E^2-\Delta^2(T))^{1/2}}$ is the quasiparticle
density of states. The most appropriate superconducting gap
$\Delta(0) =1.86k_bT_c$ is chosen in our BCS calculation with
$T_c=6.9K$, and this value is reasonable for $MgCNi_3$ because the
generally reported results are larger than the conventional BCS
value($1.76k_bT_c$). It is found that
$\lambda^{2}(0)/\lambda^{2}(T)$ of $MgCNi_3$ can be well described
by the s-wave BCS theory with a single gap, but the two-fluid
model shows a substantial deviation. This  suggests the s-wave
nature of superconductivity in $MgCNi_3$, which is consistent with
our previous conclusion reached by
point-contact-tunneling\cite{shan-tunneling}. Later on we will
show that our results are not compatible with any other pairing
symmetry with nodes on the gap function which normally contributes
a  power law dependence to the temperature dependence $\rho_s$.

For the sake of comparison, the temperature dependence of the
normalized superfluid density in $MgB_2$ obtained by exactly the
same magnetization method\cite{li-MgB2} is also shown in Fig. 5,
with  $\lambda(0)$ as a fit parameter. Clearly the data can not be
understood in isotropic s-wave BCS theory or two-fluid model
because of the two-gap characteristic of
$MgB_2$\cite{choi-natue,kang-cond}. The data obtained from this
simple magnetization method on $MgB_2$ was found to be close to
that determined by more elegant microwave
method\cite{klein-appliedSC} and radio frequency
technique\cite{manzano-prl}, which can be seen from Fig. 5. This
indicates that the same magnetization method used in $MgCNi_3$ to
get $H_{c1}$ and $\lambda$ is  reliable, and the corresponding
results are plausible.

One may argue  whether the $H_{c1}$ obtained here is the lower
critical field of grains because of the polycrystalline nature of
our sample. In our magnetization experiment, the nominal
$H_{c1}(0)$ of our sample is about 145.1 Oe. Combined with
$H_{c2}(0)$ ( $1.18\times 10^{5}$ Oe ) determined from our
previous measurement of specific-heat\cite{shan-carbon}, we can
reach that the value of $\kappa$ is 39 and $H_{c}(0)$ equals to
2165 Oe by Eqs. (3, 4). The value of coherence length is 5.3 nm
obtained by Eq. (5).
\begin{equation}
H_{c1}(T)=\frac{1}{\sqrt{2}}H_c(T)\frac{1}{\kappa}\ln\kappa
\label{eq3}
\end{equation}
\begin{equation}
H_{c2}(T)={\sqrt{2}}H_c(T)\kappa \label{eq4}
\end{equation}
\begin{equation}
\xi(0)={\sqrt{\Phi_{0}/2\pi{H_{c2}(0)}}} \label{eq5}
\end{equation}
And the value of $\lambda(0)$ is about 200.1 nm. All these values
of parameters are in the range of  the reported results of
$MgCNi_3$ by other techniques( see collected parameters in
Ref.[\onlinecite{walte}] ). This manifests that $H_{c1}$ measured
here reflects the bulk property. In addition, the value of $\xi$
for $MgCNi_3$ is quite large, so that the influence of the grain
boundary is weak.

Another argument is that the nominal $H_{c1}$ relation obtained in
our experiment may not reflect the true $H_{c1}$ but the flux
entry field because of the Bean-Livingston surface barrier and
effects of sample corners geometrical barriers. However, we would
argue that the influence of surface barrier is not important to
our cylindrical sample, since the magnetization hysteresis loops
are very symmetric in the temperature and filed regimes we
measured. In order to further verify the validity of this method
to obtain $H_{c1}$, we have repeated the same measurement for an
ellipsoidal sample. The data and the  discussion are presented
 below.

\subsection{The ellipsoidal sample ( $S$-$e$ )}

The curves of dc magnetization  are shown in Fig. 6 and the
temperature varies between  1.59 K and 6.90 K with steps about 0.1
K .  The optimal \textquotedblleft Meissner line\textquotedblright
 ( solid line in Fig.6 ) has been determined in the same way as
 for the cylindrical sample. Subtracting this ML from the
 magnetization data yields the $\Delta
M$ curves plotted in Fig. 7. The H$_{c1}(T)$ acquired by using
criteria of $\Delta M = 3.2 \times 10^{-4}$ and $1.0 \times
10^{-3}$ emu are shown in Fig. 8, and the demagnetization factor n
( $\approx0.136$ ) has been taken into account. Then the
penetration depth $\lambda(T)$ can be achieved from Eq(1) and they
are displayed in the inset of Fig. 8. The normalized temperature
dependence of $\lambda^{2}(0)/\lambda^{2}(T)$ of $MgCNi_3$ is
shown in  Fig. 9. One can clearly see that the data from the
ellipsoidal sample is almost identical to that for the cylindrical
sample, showing a trivial influence of either the geometrical or
surface barrier in our present samples.

\begin{figure}
       \centering
       \includegraphics[width=8cm]{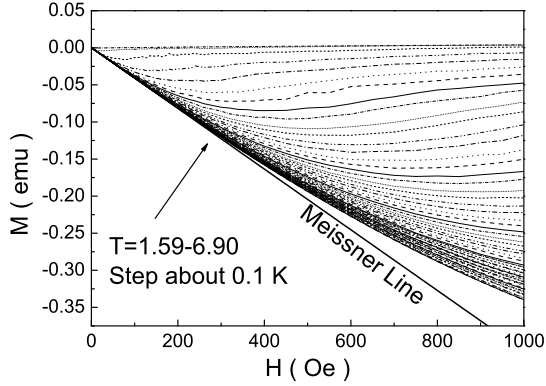}
       \caption{The magnetization curves of $M(H)$ for the ellipsoidal $MgCNi_3$ sample. The solid line is the \textquotedblleft Meissner line\textquotedblright.
       The temperature is varied from 1.59 K to 6.90 K ( from  bottom to top ),
       with  step about 0.1 K . }
       \label{figure6}
\end{figure}

\begin{figure}
       \centering
       \includegraphics[width=8cm]{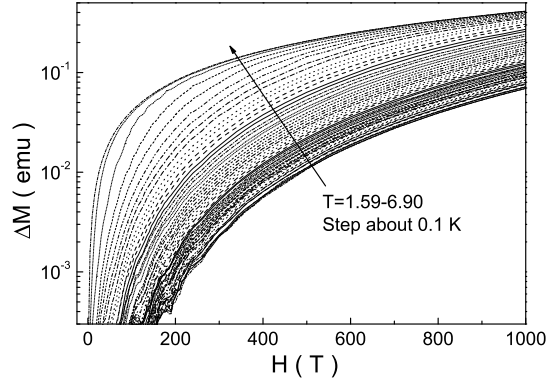}
       \caption{The difference between the \textquotedblleft Meissner line\textquotedblright~and the $M(H)$ curves. $\Delta M$ is shown in
       logarithmic scale.}
       \label{figure7}
\end{figure}

\begin{figure}
       \centering
       \includegraphics[width=8cm]{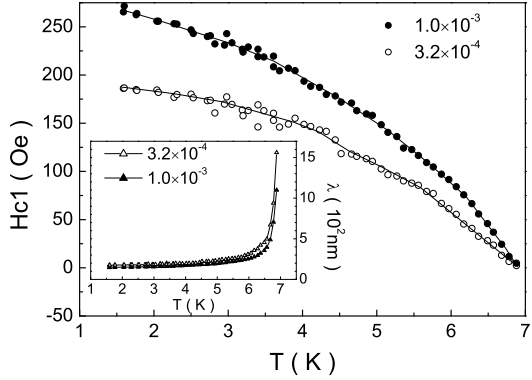}
       \caption{The temperature dependence of the  nominal $H_{c1}$ and $\lambda$ (inset).
        Lines are guides to the eyes.}

       \label{figure8}
\end{figure}

\begin{figure}
       \centering
       \includegraphics[width=8cm]{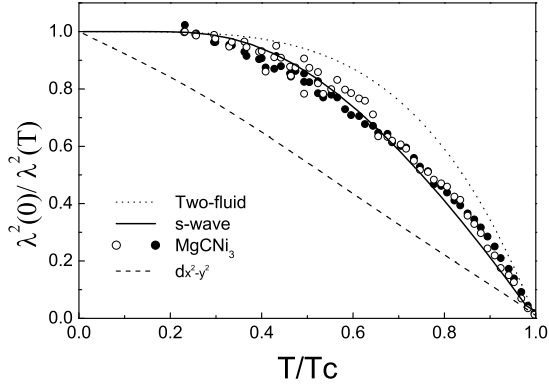}
       \caption{The temperature dependence of $\lambda^2(0)/\lambda^2(T)$ for ellipsoidal $MgCNi_3$ sample.
       The values of $\lambda^2(0)/\lambda^2(T)$ are obtained by $H_{c1}(T)$ using criteria of $\Delta M$ equal
       $3.2\times 10^{-4}$ ( open circles ) and $1.0\times 10^{-3}$ emu ( solid circles ).
       The  prediction by two-fluid model ( dotted ), BCS s-wave ( solid ) and pure d-wave ( dashed ) are also shown. }
       \label{figure9}
\end{figure}

In addition, we have calculated the  superfluid density assuming a
nodal gap with  d-wave symmetry. Under the frame of the BCS
theory, if the gap has a d-wave-like node, the normalized
superfluid density $\lambda^{2}(0)/\lambda^{2}(T)$ is written as
\begin{equation}
\frac{\lambda^{2}(0)}{\lambda^{2}(T)}=1-\frac{1}{\pi}\int^{2\pi}_{0}\int^{\infty}_{\Delta(T,\theta)}{-\frac{\partial{f(E)}
 }{\partial{E}} D(E)} \mathrm{d}E\mathrm{d}\theta
\end{equation}
with $f(E)=1/[exp(-E/{k_B}T)+1]$,
$\Delta(T,\theta)=\Delta_0(T)\cdot\cos2\theta$ and
$D(E)=E/{(E^2-\Delta^2(T,\theta))^{1/2}}$. The calculated d-wave
results are shown in  Fig. 9 with  dashed line. For
$\mathrm{p}_x$-wave symmetry with
$\Delta(T,\theta)=\Delta_0(T)\cdot\sin\theta$, we found that the
calculated temperature dependence of
$\lambda^{2}(0)/\lambda^{2}(T)$ is close to that of d-wave, and
far from our experiment data. The predictions of s-wave BCS (
solid ) and two-fluid model ( dotted ) are also shown in Fig. 9.
It is found that our data can only be  well described by the
s-wave model.  Together with the results for the cylindrical
sample, we conclude that $MgCNi_3$ is most likely an isotropic
s-wave superconductor.

\section{SUMMARY}

To summarize, we have measured the M-H curves of two $MgCNi_3$
samples with cylindrical and ellipsoidal shapes  and obtained
their lower critical field $H_{c1}(T)$ and $\lambda(T)$. The
temperature dependence of normalized superfluid density  is
consistent with the s-wave BCS theory. All these indicate that
$MgCNi_3$ may possess an isotropic s-wave gap, which is in sharp
contrast to $MgB_2$.

Note added: The recent report of carbon isotope effect in
 MgCNi$_{3}$  by T. Klimczuk and R.J. Cava indicates that carbon-based phonons play an essential role in the
superconducting mechanism \cite{Cava-prb}.

\acknowledgements This work is supported by the National Science
Foundation of China, the Ministry of Science and Technology of
China, and the Chinese Academy of Sciences within the knowledge
innovation project.


\end{document}